\documentclass[11pt,twoside]{article}

\usepackage{asp2006}
\usepackage{epsf}
\usepackage{psfig}
\usepackage{lscape}

\begin{document}
\title{Probing Quasar Outflows with Intrinsic Narrow Absorption Lines} 

\author{Toru Misawa\altaffilmark{1}, Michael
Eracleous\altaffilmark{1,2,3}, Jane C. Charlton\altaffilmark{1}, Rajib
Ganguly\altaffilmark{4}, David Tytler\altaffilmark{5,6}, David
Kirkman\altaffilmark{5,6}, Nao Suzuki\altaffilmark{5,6}, and Dan
Lubin\altaffilmark{5,6}}

\altaffiltext{1}{Department of Astronomy \& Astrophysics, The
  Pennsylvania State University, University Park, PA 16802}
\altaffiltext{2}{Department of Physics \& Astronomy, Northwestern
  University, 2131 Tech Drive, Evanston, IL 60208}
\altaffiltext{3}{Center for Gravitational Wave Physics, The
  Pennsylvania State University}
\altaffiltext{4}{Department of Physics and Astronomy, 1000 East
  University Ave, University of Wyoming (Dept 3905), Laramie, WY,
  82071}
\altaffiltext{5}{Center for Astrophysics and Space Sciences,
  University of California San Diego, MS 0424, La Jolla, CA
  92093-0424}
\altaffiltext{6}{Visiting Astronomer, W. M. Keck Observatory, which is
  a joint facility of the University of California, the California
  Institute of Technology, and NASA}

\begin{abstract} 

We present statistical and monitoring results of narrow absorption
lines that are physically related to quasars (i.e., {\it intrinsic}
NALs).
We use Keck/HIRES spectra of 37 optically bright quasars at $z$=2--4,
and identify 150 NAL systems that contain 124 {\rm C}~{\sc iv}, 12
{\rm N}~{\sc v}, and 50 {\rm Si}~{\sc iv} doublets. Among them, 39 are
classified as intrinsic systems based on partial coverage analysis. At
least 50\% of quasars host intrinsic NALs. We identify two families of
intrinsic systems based on their ionization state. Some intrinsic
systems have detectable low-ionization NALs at similar velocities as
higher-ionization NALs, although such low-ionization lines are rare in
broad absorption line (BAL) systems.
We also have observed an optically bright quasar, HS1603+3820, eight
times with Subaru/HDS and HET/MRS over an interval of 4.2 years (1.2
years in the quasar rest frame), for the purpose of monitoring a
variable {\rm C}~{\sc iv} mini-BAL system. We find that all the
troughs of the system vary in concert. However, no other correlations
are seen between the variations of different profile parameters. We
propose that the observed variations are either (i) a result of rapid
continuum fluctuations, caused by a clumpy screen of variable optical
depth located between the continuum source and the mini-BAL gas, or
(ii) a result of variable scattering of continuum photons around the
absorber.

\end{abstract}

\section{Introduction}

Outflowing winds from the accretion disks around supermassive black
holes provide a potential mechanism for extracting angular momentum
from the accreting material, allowing accretion to proceed. They are
also cosmologically important because they provide energy and momentum
feedback to the intergalactic medium (IGM). Traditionally, broad
absorption lines (BALs; FWHM $\geq$ 2,000 km~s$^{-1}$) have been used
to study such outflowing winds, because their large line widths and
smooth line profiles suggest an association with the wind. Intrinsic
narrow absorption lines (NALs; FWHM $\leq$ 500 km~s$^{-1}$) that are
physically related to the quasars are an alternative and more
promising tool for examining the physical conditions of the outflow
than BALs for two important reasons: (i) NALs do not suffer from
self-blending (i.e., a blend of blue and red members of doublets such
as {\rm C}~{\sc iv}$\lambda\lambda$1548,1551), and (ii) NALs are found
in a wider variety of AGNs, while BALs are detected primarily in
radio-quiet quasars.

In spite of their potential importance, NALs have not received as much
attention as BALs, because it is difficult to distinguish intrinsic
NALs from NALs that are not physically related to the quasars (i.e.,
{\it intervening} NALs), produced in intervening galaxies, the IGM,
Milky Way gas, or gas in the quasar host galaxies. With the advent of
high-dispersion spectroscopy of faint objects, it has become possible
to identify intrinsic NALs, based primarily on one or both of the
following indicators: (a) the dilution of absorption troughs by
unocculted light (e.g., Hamann et al. 1997), and (b) time variability
of line profiles (e.g., depth, equivalent width, and centroid), within
a year in the absorber's rest frame (e.g., Barlow \& Sargent 1997).

\section{Statistical Analysis with 37 Keck/HIRES Quasars}

We have constructed a large, relatively unbiased, equivalent width
limited sample of {\it intrinsic} NAL systems found in the
high-resolutional ($R$ = 37,500) spectra of $z$=2--4 quasars taken
with Keck/HIRES. We identify 150 NAL systems that contain 124 {\rm
C}~{\sc iv}, 12 {\rm N}~{\sc v}, and 50 {\rm Si}~{\sc iv} doublets, of
which 39 are identified as intrinsic NAL systems on the basis of their
partial coverage signature. Using this sample, we study their
demographics, the distribution of their physical properties, and any
relations between them.

\subsection{Fraction of Intrinsic NALs and Quasars with Intrinsic NALs}

We separate intrinsic and intervning NAL systems using the partial
coverage (PC) of doublets (e.g., Wampler, Chugai, \& Petitjean
1995). Of the 124 {\rm C}~{\sc iv} doublet, we find 19\%\ show
PC. Similarly, 18\%\ of {\rm Si}~{\sc iv} doublets and 75\%\ of {\rm
N}~{\sc v} doublets show PC. If we focus on only systems near the
quasar redshift ($v_{shift}$ $<$ 5000 km~s$^{-1}$; hereafter
associated NALs), the fraction of {\rm C}~{\sc iv} doublets showing PC
increases to 33\%. No associated {\rm Si}~{\sc iv} doublet shows PC,
while all {\rm N}~{\sc v} doublets showing PC are associated. (For
{\rm N}~{\sc v}, we can only probe associated systems due to severe
Ly$\alpha$ forest contamination at larger velocities). Richards (2001)
estimate that as many as $\sim$36\%\ of {\rm C}~{\sc iv} NALs may be
intrinsic systems appearing at high ejection velocity. Taken at face
value, this may imply that only $\sim$50\%\ of intrinsic systems show
PC.

We also find that the fraction of quasars that have one or more
intrinsic NAL systems is about 50\%, although this is a lower limit
because our spectra do not have full offset velocity coverage and
because some intrinsic absorbers may not exhibit the signature of
partial coverage. One possible interpretation of this result is that
50\%\ of the solid angle around a typical quasar is covered by the
absorbers, {\it on average}.

\subsection{Ionization Conditions of Intrinsic NAL Systems}

Considering the ionization structure of the 39 intrinsic NAL systems
in our sample, we find the following two major categories, which may
represent absorbers of different densities and/or at different
distances from the ionizing source.

\begin{description}

\item[Strong {\rm C}~{\sc iv} Systems] --- These are characterized by
 strong (i.e., large equivalent width), partially covered {\rm C}~{\sc
 iv} NALs, strong Ly$\alpha$ lines, and relatively weak or
 undetectable {\rm N}~{\sc v} NALs. Of the 28 systems in this
 category, 25 have intermediate ionization lines (such as {\rm
 Si}~{\sc iii} and {\rm C}~{\sc iii}) and 15 have low ionization
 absorption lines (such as {\rm Si}~{\sc ii} and {\rm C}~{\sc ii})
 (see Figure~1).

\item[Strong {\rm N}~{\sc v} Systems] --- These are characterized by
 strong {\rm N}~{\sc v} NALs, and relatively weak, non-black
 Ly$\alpha$ lines. {\rm C}~{\sc iv} and {\rm O}~{\sc vi} NALs may also
 be detected in these systems, and in some cases {\rm O}~{\sc vi} may
 be stronger than {\rm N}~{\sc v}. We find 11 systems in this
 category, of which 5 have detected intermediate ionization
 transitions, and only 1 has detected low ionization transitions (see
 Figure~1).

\end{description}

\begin{figure}
  \plottwo{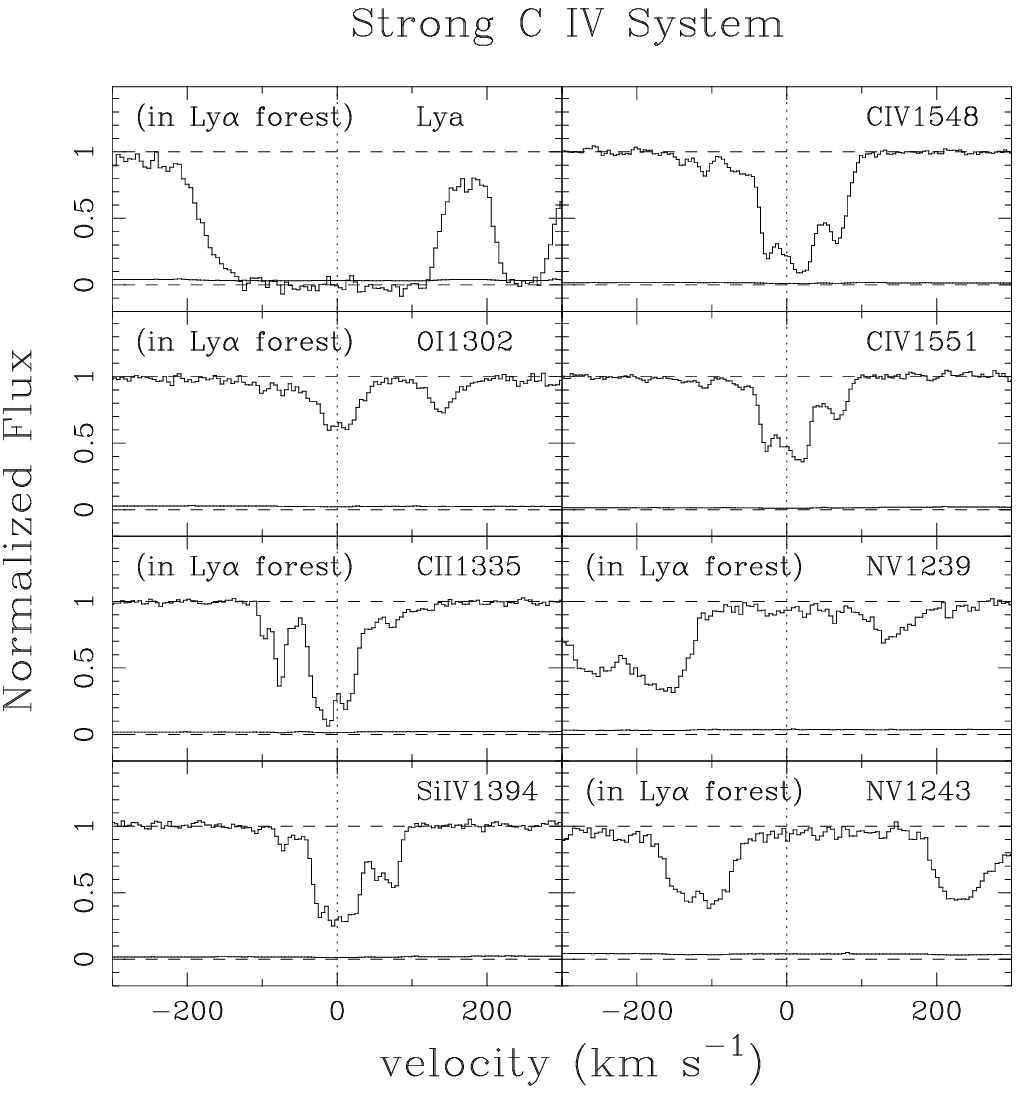}{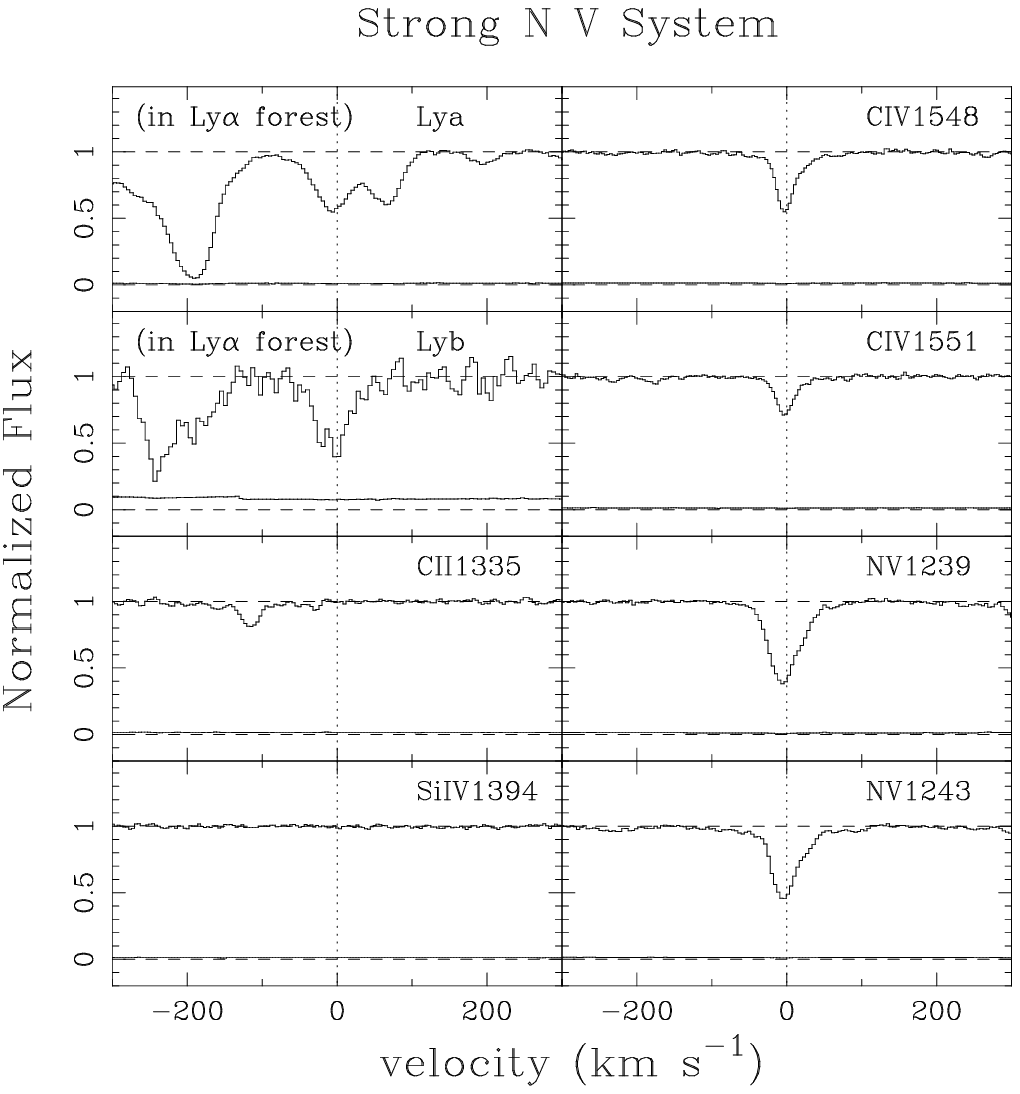}
  \caption{Velocity plots of various transitions in the system at
  $z_{abs}$ = 1.8919 of HS1103+6416 (strong {\rm C}~{\sc iv} system)
  and the system at $z_{abs}$ = 2.7125 of HS1700+6416 (strong {\rm
  N}~{\sc v} system). We classify 39 intrinsic NAL systems into 28
  strong {\rm C}~{\sc iv} and 11 strong {\rm N}~{\sc v} systems.}
\end{figure}

In the 15 strong {\rm C}~{\sc iv} systems with low ionization
transitions detected, the line profiles of the {\rm C}~{\sc iv} NALs
and the low ionization lines are similar. Moreover, the low ionization
lines are usually detected at the same velocities as the partially
covered {\rm C}~{\sc iv} NALs, implying that both families of lines
arise in the same parcels of gas. About 50\%\ of intrinsic NAL systems
include low-ionization lines, while the fraction of BAL systems with
low-ionization lines (i.e., LoBALs) among all BAL systems is only
13--17\%\ (e.g., Sprayberry \& Foltz 1992; Reichard et al. 2003). The
different ionization conditions of intrinsic NALs and BALs suggest
that they may be located in different regions around the accretion
disk.

\section{Monitoring of C IV Mini-BAL in HS1603+3820}

We have obtained six Subaru/HDS and two HET/MRS spectra of the quasar
HS1603+3820 ($z_{em}$ = 2.542) spanning an interval of approximately
1.2 years in the quasar rest frame, for the purpose of monitoring
intrinsic systems. Among 9 {\rm C}~{\sc iv} systems, only the mini-BAL
system at $z_{abs}$ $\sim$ 2.43 ($v_{shift}$ $\sim$ 9,500 km~s$^{-1}$
from the quasar emission redshift) shows both time variability and
partial coverage.
We fit models only to the bluest portion of the {\rm C}~{\sc iv}
mini-BAL profile where self-blending is not severe. All fit parameters
(i.e., column density, Doppler parameter, coverage fraction, and shift
velocity) as well as the total equivalent width of the system vary
significantly with time, even on short time scales. However, the
profile parameters do not correlate with each other, with one
exception: the equivalent widths of all the troughs in this system
vary together.
We have examined a number of ways of explaining the above variations
of the {\rm C}~{\sc iv} mini-BAL and we have found two viable
possibilities.

\begin{description}

\item[Scattering of Continuum Photons] --- The observed partial
 coverage signature is the result of continuum photons scattering
 around the absorber and into our cylinder of sight. The observed
 changes in mini-BAL equivalent widths are thus produced by variations
 in the scattered continuum that dilutes the absorption troughs.  This
 idea can be tested observationally through spectropolarimetry (e.g.,
 Brotherton et al. 1997).

\item[Screening of Variable Optical Depth] --- The illumination of the
 UV absorber fluctuates on short time scales. We suggest that these
 fluctuations are caused by a screen of variable optical depth between
 the mini-BAL gas and the continuum source. This screen might be
 identified with the shielding gas invoked or predicted in some
 outflow models.  Moreover, it could be analogous to the ``warm''
 absorbers observed in the X-ray spectra of Seyfert galaxies and some
 quasars (Crenshaw et al. 1999). This picture can also explain the
 variations in the coverage fraction, which appear to be unrelated to
 the ionic column density changes.

\end{description}

\acknowledgements

This work was supported by NASA grant NAG5-10817. UCSD work was
supported in part by NASA grant NAG5-13113 and NSF grant AST 0507717.
ME acknowledges partial support from the Theoretical Astrophysics
Visitors' Fund at Northwestern University.

%%% THE BIBLIOGRAPHY

\end{document}